\documentclass{amsart}
\usepackage[latin1]{inputenc}
\usepackage{amsmath}
\usepackage{amsfonts}
\usepackage{amssymb}
\newcommand{\mf}{\mathfrak}
\newcommand{\mc}{\mathcal}
\newcommand{\be}{\begin{equation*}}
\newcommand{\ee}{\end{equation*}}
\newcommand{\orb}{\mc{O}(a)}

\begin{document}

\title{On geometric quantisation of the Dirac magnetic monopole}

\author{G.M. Kemp}
\address{Department of
Mathematical Sciences,
          Loughborough University,
Loughborough,
          Leicestershire, LE11 3TU, UK}
       \email{G.Kemp@lboro.ac.uk}

\author{A.P. Veselov}
\address{Department of
Mathematical Sciences,
          Loughborough University,
Loughborough,
          Leicestershire, LE11 3TU, UK
          and
Moscow State University, Moscow 119899, Russia}

\email{A.P.Veselov@lboro.ac.uk}

\maketitle

{\small  {\bf Abstract.} We give a simple derivation of the spectrum of the Dirac magnetic monopole on a unit sphere $S^2$ based on geometric quantization and the Frobenius reciprocity formula.  We also briefly discuss the generalisations of Dirac magnetic monopole to any coadjoint orbit of a compact Lie group.}

\section{Introduction}
The Dirac magnetic monopole is one of the most remarkable and one of the first integrable systems of quantum mechanics. In his pioneering paper \cite{Dirac} Dirac showed that an isolated magnetic charge $q$ should be quantized: $q \in \mathbb Z.$  
The corresponding Schr\"odinger equation was solved by Tamm \cite{Tamm} while he was visiting Dirac in Cambridge in 1931.

On the other hand it took quite a while to understand the global nature of the corresponding eigenfunctions. For the Dirac magnetic monopole on a sphere it was done only in 1976 by
Wu and Yang \cite{WY}, who explained that the corresponding wavefunctions (known as {\it monopole harmonics}) are sections of a complex line bundle $L$ over $S^2$ and found the spectrum to be 
\begin{equation}
\lambda = \left[l(l+1)+|q|\left(l+\frac{1}{2}\right)\right], l=0,1,2,\dots \ \text{with degeneracy } \  2l+|q|+1. \label{eq:spec}
\end{equation}
This gives also a geometric interpretation of Dirac's quantization condition: magnetic charge is the first Chern class $c_1(L)=q$ of the bundle $L$, which must be an integer.  

A different derivation of this result was given by Ferapontov and one of the authors  \cite{FV}, who extended  the classical factorisation method going back to Darboux and Schr\"odinger \cite{Sch} to curved surfaces. This provided an explicit description of the monopole harmonics by recursive application of the lowering operators to the ground states, which, under the isomorphism $S^2 \cong \mathbb{C}P^1,$  are given for positive $q$ by polynomials of degree $\leq q.$

We present here a simple derivation of the spectrum of the Dirac monopole on a unit sphere using geometric quantization. We should say that geometric quantization of the Dirac magnetic monopole and related problems were already discussed in \cite{Mlad, Prieto}, but we believe that our approach is simpler and clearer.

The initial point for us was the calculation by Novikov and Schmelzer \cite{NS} of the canonical symplectic structure on the coadjoint orbits of the Euclidean group $E(3)$ of motions  of $\mathbb{E}^3,$ which showed the relation with the classical Dirac monopole (see the next section). A similar calculation for Poincare and Galilean groups was done by Reiman \cite{Rei}, who also seems to have the idea of geometric quantization in mind, but did not pursue it. 

We first show that Novikov-Schmelzer variables have a natural quantum version as covariant derivatives acting on the space of sections $\Gamma(L)$ of the corresponding line bundle $L.$ 
Fierz's modification \cite{Fierz} of the angular momentum in the presence of the Dirac magnetic monopole appears naturally in this relation.

The space $\Gamma(L)$ is the representation of $SU(2)$  induced from the representation of $U(1) \subset SU(2)$ given by $z \rightarrow z^{q}, z \in U(1).$ This space can be decomposed as an $SU(2)$-module using  the classical Frobenius reciprocity formula \cite{FH}.
We show that the formula for the Dirac monopole spectrum (\ref{eq:spec}) is a simple corollary of this.

In the last section we discuss the generalisations of this to any coadjoint orbit of a compact Lie group.

\section{The coadjoint orbits of the Euclidean group $E(3)$}

Let $e(3)$ be the Lie algebra  of the Euclidean group $E(3)$ of motions of $\mathbb{E}^3.$ It has the basis $l_1,l_2,l_3,p_1,p_2,p_3$, where $p$ and $l$ are generators of translations and rotations (momentum and angular momentum) respectively. 

The dual space $e(3)^*$ with the coordinates $\left\{l_1,l_2,l_3,p_1,p_2,p_3\right\}$ has the canonical Poisson bracket
\begin{equation*}
\left\{l_i, l_j\right\} = \epsilon_{ijk} l_k, \ \  \left\{l_i, p_j\right\} = \epsilon_{ijk} p_k, \ \  \left\{p_i, p_j \right\}=0.
\end{equation*}
Its symplectic leaves are the coadjoint orbits of $E(3),$ which are the level sets of the Casimir functions
\begin{equation*}
C_1:=(p,p)=R^2, \qquad C_2:=(l,p)=\alpha R
\end{equation*} 
Following Novikov and Schmelzer \cite{NS} introduce
\begin{equation}
\label{NS}
 \sigma_i=l_i-\frac{\alpha}{R}p_i
 \end{equation}
 to identify the coadjoint orbits with $T^*S^2:$ 
\begin{equation*}
(p,p)=R^2, \qquad (\sigma,p)=0
\end{equation*}
where $TS^2$ and $T^*S^2$ have been identified using the standard Riemannian metric on the radius $R$ sphere.

The new coordinates $\left\{\sigma_1,\sigma_2,\sigma_3,p_1,p_2,p_3\right\}$ have Poisson brackets
\begin{equation}
\left\{\sigma_i, \sigma_j\right\} = \epsilon_{ijk}\left(\sigma_k - \frac{\alpha}{R}p_k \right), \ \  \left\{\sigma_i, p_j\right\} = \epsilon_{ijk} p_k, \ \  \left\{p_i, p_j \right\}=0. \label{eq:PBs}
\end{equation}

Novikov and Schmelzer computed the canonical symplectic form on the coadjoint orbits and showed that it is given by
\begin{equation}
\omega=\, \mathrm{d} P \wedge \, \mathrm{d} Q +  \frac{\alpha}{R^2}  \, \mathrm{d} S \label{eq:symp}
\end{equation}
where $\, \mathrm{d} P \wedge \, \mathrm{d} Q$ is the standard symplectic form on $T^* S^2$ and $\, \mathrm{d} S$ is the area form on $S^2$ (see also \cite{Rei}). As it was pointed out in \cite{NS} the second term corresponds to the magnetic field of the (non-quantized) Dirac monopole:
\begin{equation*}
\mathcal H= \frac{\alpha}{R^2} \, \mathrm{d} S.
\end{equation*} 

The value $$q= \frac{1}{2\pi} \int_{S^2} \mathcal H$$ is called the {\it charge} of the Dirac monopole.
Dirac's {\it quantization condition} \cite{Dirac} is
$$q=\frac{1}{2\pi} \int_{S^2} \mathcal H = \frac{1}{2\pi} \int_{S^2} \frac{\alpha}{R^2} \mathrm{d} S=2 \alpha \in \mathbf Z.$$
Comparing this with (\ref{eq:symp}) 
we see that this is identical to the {\it geometric quantization} condition \cite{Kir}:
$$\frac{1}{2\pi} \int_{[S^2]} \omega =\frac{1}{2\pi} \int_{S^2} \frac{\alpha}{R^2} \mathrm{d} S \in \mathbf Z.$$

\section{Line bundles over $S^2$ and quantization of Novikov-Schmelzer variables}

It is convenient to use the scaled variables
\begin{equation}
x_i = p_i/R, \quad x^2=x_1^2+x_2^2+x_3^2=1
\label{scaled}
\end{equation}
to work with the unit sphere $S^2.$

The quantum version of the Poisson brackets (\ref{eq:PBs}) are the following commutation relations
(we are using the units in which Planck's constant $\hbar=1$)
\begin{equation}
[\hat\sigma_k, \hat\sigma_l] = i\epsilon_{klm}(\hat\sigma_m - \alpha \hat x_m), \ \  [\hat\sigma_k, \hat x_l] = i \epsilon_{klm} \hat x_m, \ \  [\hat x_k, \hat x_l]=0. \label{eq:qPBs}
\end{equation}

We are going to show now that this algebra has a natural representation on the space of sections of a certain line bundle over $S^2.$ 

Recall that a connection on a vector bundle $\xi$ over manifold $M^n$ associates to every vector field $X$ on $M^n$ the operator of covariant derivative $\nabla_X$ acting on sections of $\xi.$ The corresponding curvature tensor $\mathcal R$ is defined for each pair of vector fields $X,Y$ as
\begin{equation*}
\mathcal R(X,Y)=\nabla_X \nabla_Y - \nabla_Y \nabla_X  - \nabla_{[X,Y]},
\end{equation*}
where $[X,Y]$ is the standard Lie bracket of vector fields (see e.g. \cite{KN}).

Consider a complex line bundle over $S^2$ with a $U(1)$-connection having the curvature form
\begin{equation*}
\mathcal  R= i \mathcal  H = i \alpha \, \mathrm{d} S,
\end{equation*} 
which is motivated by geometric quantization.
Since the first Chern class of the bundle must be an integer we have 
$$q=\frac{1}{2\pi i} \int_{S^2} \mathcal R = \frac{1}{2\pi} \int_{S^2} \alpha \mathrm{d} S=2 \alpha \in \mathbf Z,$$
which is precisely Dirac's quantization condition.

 Let $$X_1=x_3 \partial_2 - x_2 \partial_3,\ X_2= x_1 \partial_3 - x_3 \partial_1,\ X_3= x_2 \partial_1 - x_1 \partial_2$$ be the vector fields generating rotations of $S^2$ and $\nabla_{X_j}$ be the corresponding covariant derivatives. 
We claim that 
$$\hat \nabla_j:=i \nabla_{X_j}$$ and the operators $\hat x_j$ of multiplication by  $x_j$ satisfy the commutation relations (\ref{eq:qPBs}).

Indeed, by definition of the curvature form, we have
$$
\mathcal  R(X_1, X_2)=\nabla_{X_1} \nabla_{X_2} - \nabla_{X_2} \nabla_{X_1} - \nabla_{[X_1,X_2]} = i \alpha x_3
$$
since
$$
\alpha \, \mathrm{d} S (X_1,X_2)= \alpha \left| \begin{array}{ccc}
x_1 & x_2& x_3  \\
0 & x_3 & -x_2 \\
-x_3 & 0 & x_1\\
\end{array} 
\right| \nonumber \\
= \alpha x_3(x_1^2 + x_2^2 + x_3^2) =  \alpha x_3.  
$$
This implies
$$[\nabla_{X_1}, \nabla_{X_2}]=\nabla_{X_3}+i \alpha \hat x_3$$
since $[X_1,X_2]=X_3.$ Similarly we have 
$$[\hat \nabla_k, \hat \nabla_l] = i \epsilon_{klm}(\hat \nabla_m - \alpha \hat x_m)$$
for all $k,l,m=1,2,3.$ The rest of the relations (\ref{eq:qPBs}) are obvious.

Alternatively, we can look for the quantization of Novikov-Schmelzer variables as covariant derivatives: $$\hat\sigma_j=i \nabla_{X_j}.$$
Then the same calculation shows that the curvature form of the corresponding connection must be 
$ i \alpha \, \mathrm{d} S.$ 

Finally returning to the original variables we have the operators
\begin{equation}
\hat l_j = \hat \nabla_j + \alpha x_j,
\label{mam}
\end{equation}
which coincides with the famous modification of the angular momentum in the presence of the Dirac magnetic monopole \cite{Fierz}. This provides us with one more explanation of this well known, but a bit mysterious \footnote {Sidney Coleman, in his famous lectures on Dirac monopoles \cite{Col}, wrote about this modification of angular momentum:``The second term looks very strange indeed; in Rabi's immortal words about something else altogether,``Who ordered that ?""} physical notion.

\section{Induced representations and Frobenius reciprocity }

Let $L_q$ be the complex line bundle over $S^2$ with first Chern class $q$. 
We are interested in  the space $\Gamma(L_q)$ of sections of $L_q$.  Viewing $S^2$ as $SU(2)/U(1)$ (with $U(1)$ as the diagonal subgroup) we have a natural interpretation of $\Gamma(L_q)$ as a representation of $SU(2).$ 

In representation theory this construction is known as an induced representation (see e.g. \cite{FH}).
One can use the classical Frobenius reciprocity formula from this theory to decompose $\Gamma(L_q)$ into irreducible $SU(2)$ modules. 

First recall that all finite-dimensional irreducible representations of $SU(2)$ are labelled by a highest weight $k \in \mathbf Z_{\geq 0}$. The corresponding spaces $V_k$ have  dimension $k+1$ and weights
\begin{equation}
-k, -k+2, \ldots, k-2, k. \label{eq:weights}
\end{equation}
All finite-dimensional irreducible representations $W_q$ of $U(1)$ have dimension 1 and are given by 
\begin{equation*}
e^{i \theta} \mapsto e^{i q \theta}, \,\, q \in \mathbf Z.
\end{equation*}

One can use $W_q$ to induce a representation  $\mathrm{ind}_{U(1)}^{SU(2)}\left(W_q\right)$  of $SU(2)$, which can be described geometrically as the space of sections of the line bundle $L_q$ over $S^2$ with the first Chern class $q$ (see e.g. \cite{Bott}):
\begin{equation*}
\Gamma(L_q)=\mathrm{ind}_{U(1)}^{SU(2)}\left(W_q\right).
\end{equation*}

This induced representation is not irreducible. To decompose it we will use the {\it Frobenius reciprocity formula} 
\begin{equation}
\left\langle V,\mathrm{ind}_{H}^{G}\left(W\right) \right\rangle_{G}=\left\langle W, \mathrm{res}_{G}^{H}(V)\right\rangle_{H}. \label{eq:frobG}
\end{equation}
Here $G$ is a group, $H$ is its subgroup, $V$ and $W$ are the irreducible representations of $G$ and $H$ respectively, $\mathrm{ind}_{H}^{G}(W)$ is the representation of $G$ induced from $W$,
$\mathrm{res}_{G}^{H}(V)$ is the restriction of the representation $V$ to the subgroup $H$ and the brackets denote the multiplicity of the first representation entering into the second one (see e.g. \cite{FH}).

In our concrete case we have
\begin{equation}
\left\langle V_k,\mathrm{ind}_{U(1)}^{SU(2)}\left(W_q\right) \right\rangle_{SU(2)}=\left\langle W_q, \mathrm{res}_{SU(2)}^{U(1)}(V_k)\right\rangle_{U(1)}. \label{eq:frob}
\end{equation}

Since the restriction is the sum of the weight spaces $$\mathrm{res}_{SU(2)}^{U(1)}(V_k)=\bigoplus_{j \in S_k} W_j,$$ where 
$S_k= \{-k, -k+2, \ldots, k-2,k\}$ we see that each $V_k,$ which (after restriction) contains $W_q$ will appear once  in the decomposition of $\Gamma(L_q)$ and this can happen only if $k\geq |q|$ and $k-|q|$ is even.   Therefore $\Gamma(L_q)$  decomposes into $SU(2)$-modules as 
\begin{equation}
\mathrm{ind}_{U(1)}^{SU(2)}\left(W_q\right) = \Gamma(L_q)=\bigoplus_{l\in\mathbb{Z}_{\geq0}}{V_{2l+|q|}}. \label{eq:decomp}
\end{equation}

\section{Calculation of the spectrum of the Dirac monopole}

The Hamiltonian of the Dirac monopole can be written in terms of Novikov-Schmelzer operators as
\begin{equation*}
H=\hat \sigma^2
\end{equation*}  
or, equivalently,  in terms of magnetic angular momentum $\hat l$ as
\begin{equation*}
H=\hat l^2-\alpha^2=\hat l^2-\frac{1}{4} q^2.
\end{equation*}
Since the components of $\hat l_m$ satisfy the standard commutation relations
\begin{equation*}
[\hat l_k, \hat l_m] = i\epsilon_{kmn}\hat l_n,
\end{equation*}
the operator $\hat l^2$ is a Casimir operator for $SU(2)$ and 
 acts on $V_k$ as a scalar: if $s=k/2$ then 
\begin{equation}
\hat l^2 =  s(s+1) = \frac{1}{4}k(k+2), \label{eq:Lsquared}
\end{equation} 
see e.g. \cite{FH}. 
The space $V_{2l+|q|}$ has dimension $2l + |q| +1$, and for $\psi \in V_{2l+|q|}$, the operator $H$ acts as
$$
H\psi=(\hat l^2-\frac{1}{4} q^2)\psi=\left[\frac{1}{4}\left(2l+|q|\right) \left(2l+|q|+2\right) - \frac{1}{4} q^2 \right]\psi=  \left[l(l+1)+|q|\left(l+\frac{1}{2}\right) \right]\psi
$$
Thus for a Dirac monopole of charge $q$ the spectrum is  
\begin{gather*}
 \left[l(l+1)+|q|\left(l+\frac{1}{2}\right)\right], l\in\mathbb{Z}_{\geq0} \ 
\text{with degeneracy } \  2l+|q|+1
\end{gather*}
agreeing exactly with \eqref{eq:spec}.  

The corresponding eigenfunctions were computed using Darboux-Schr\"odinger factorisation method in \cite{FV}, where the ground eigenstates were identified
for non-negative q with the space of polynomials of degree $\leq q$. In our picture
the ground eigenspace corresponds to the subspace of holomorphic sections of $L_q$,
which by the Borel-Weil theorem \cite{Bott, FH} can be identified with the corresponding
irreducible SU(2)-module $V_q$.

\section{Generalisation: Dirac magnetic monopole on coadjoint orbits }

We discuss briefly here the generalisation of this to any coadjoint
orbit of a compact Lie group \footnote{We are very grateful to Alexey Bolsinov for illuminating discussions of this generalisation.} referring for the details to \cite{Kemp}.
The classical case was studied by Efimov \cite{E} and Bolsinov and Jovanovic \cite{BJ}.

Let $G$ be a Lie group,
$\mf{g}$ be its Lie algebra, $\mf{g}^*$ be its dual space, $Ad$ and $Ad^*$ be the corresponding adjoint and coadjoint actions of $G$ on $\mf{g}$ and $\mf{g}^*$:
\be
Ad_g(Y)= \left.\frac{d}{ds} \right|_{s=0} g \exp(sY) g^{-1} = gYg^{-1}.
\ee  
\be
\left\langle Ad_g^*(f),X\right\rangle := \left\langle f, Ad_{g^{-1}}X\right\rangle.
\ee
where $\left\langle f,X\right\rangle$ is the canonical pairing between $f\in \mf{g}^*$ and $X \in \mf{g}$.
Similarly the adjoint and coadjoint actions $ad^*$ of $\mf{g}$ on $\mf{g}^*$ are defined by
\be
ad_X(Y) = \left.\frac{d}{dt} \right|_{t=0} \exp(tX) Y \exp(-tX) =  [X,Y],
\ee\be
\left\langle ad_Y^*(f),X\right\rangle := -\left\langle f, ad_{Y}X\right\rangle = \left\langle f, [X,Y]\right\rangle.
\ee
Given $a$ in $\mf{g}^*$, the {\it coadjoint orbit} $\orb$ is defined by
\be
\orb := \left\{x \in \mf{g}^* : x = Ad_g^*(a), g \in G\right\} = Ad_G^*(a).
\ee
Defining $G_a$ to be the stabilizer of the point $a$,  $G_a := \left\{g\in G: Ad_g^*(a)=a\right\}$, it is clear that if $x =Ad^*_g(a)$ then $G_x = g(G_a)g^{-1}$.  In this way, one may identify $\orb$ with the homogeneous space $\orb \cong G/G_a$. 

Assume now that $G$ is compact, connected, simply-connected semi-simple Lie group.
Such a group has a positive $Ad$-invariant {\it Cartan--Killing form} $(\ ,\ )$ defined on its Lie algebra $\mf{g}$ (see e.g. \cite{FH}), which can be used to identify $\mf{g}^*$ with $\mf{g}$. 

There are two natural metrics on $\mc{O}(a)$: the metric induced from $\mf{g}^*$ with Cartan--Killing form and 
 the following {\it normal metric} defined by
\be
(\xi_x,\eta_x)_{nor} := (\pi(\xi), \pi(\eta)).
\ee
where $\pi$ is the projection map $\pi: \mf{g} \rightarrow \mf{g}/\mf{g}_x \cong \mf{g}_x^\perp.$ We will be using the latter one.

Consider the following class of closed $G$-invariant 2-forms on $\orb$ (cf. \cite{BJ}). 
Let $G_x \subset G$ be the stationary subgroup of $x \in \orb$ and $Z(\mf{g}_x)$ be the centre of its Lie algebra. 
 Choose $f_a \in (Z(\mf{g}_a))^*$ and 
define $f_x = Ad^*_g(f_a) \in (Z(\mf{g}_x))^*$ at any other point $x = Ad^*_g(x) \in \orb$. 

Let $\xi_x = ad^*_\xi(x)$ and $\eta_x = ad^*_\eta(x)$ be two tangent vectors at $x.$ Define a 2-form on $T_x\orb$ by
\be
\sigma_{f}(\xi_x, \eta_x) = \left\langle f_x, [\xi,\eta] \right\rangle.
\ee
Then $\sigma_f$ defines an invariant, closed 2-form on $\orb$. When $f_a=a$ we have  a canonical, $G$-invariant {\it Kostant--Kirillov symplectic form} 
\be
\omega_{KK}(\xi_x,\eta_x) = \left\langle x, [\xi,\eta]\right\rangle.
\ee

The forms $\sigma_f$ determine the class of $G$-invariant magnetic fields on $\orb$, which can be considered as classical analogues of Dirac magnetic monopoles. The original Dirac magnetic monopole discussed above corresponds to the simplest case $G=SO(3)$ and $\orb = S^2.$
The Lie group $E(3)$ above in general case should be replaced by the semi-direct product $EG=G \ltimes_{Ad} \mf{g}$ with the product
\be
(g_1,X_1) \cdot (g_2,X_2) = (g_1g_2,Ad_{g_2^{-1}}X_1 + X_2). \label{eq:liegroup}
\ee
The Lie algebra structure on $\mf{eg}=\mf{g} \ltimes_{Ad} \mf{g}$ is given by
\be
\left[(u_1,v_1),(u_2,v_2)\right]=\left(\left[u_1,u_2\right],\left[u_1,v_2\right] - \left[u_2,v_1\right]\right). \label{eq:comms}
\ee
One can show \cite{Kemp} that the coadjoint orbit $\mc{O}(f,a)$ of a point $(f,a) \in \mf{eg}^*$ with canonical Kostant--Kirillov symplectic form is symplectomorphic to the magnetic cotangent bundle $(T^*\mc{O}(a), dp\wedge dq + \pi^*(\omega_f)),$ which is a generalisation of the Novikov--Schmelzer result. 

By the {\it generalised Dirac magnetic monopole on coadjoint orbit} we mean the system on $T^*\mc{O}(a)$ with the symplectic form $\omega=dp\wedge dq + \pi^*(\omega_f))$ and the Hamiltonian given by the normal metric on $\mc{O}(a)$.
Efimov \cite{E} and Bolsinov and Jovanovic \cite{BJ} proved the classical integrability in the case when the magnetic field is a multiple of Kirillov--Kostant form: $f=\alpha x.$

The quantisation gives the integrality condition for the corresponding form $\omega_f:$ 
$$\frac{1}{2\pi} \int_\sigma \omega_f \in \mathbb Z$$
for all $\sigma \in H_2(\mc{O}(a), \mathbb Z).$ Such forms are in one-to-one correspondence with the characters of $H=G_a$ and with homogeneous line bundles over $\mc{O}(a)=G/H.$ 

The quantum Hamiltonian of the generalised Dirac magnetic monopole is nothing else but the {\it Bochner Laplacian} (see e.g. \cite{Prieto} and references therein) acting in the sections of the corresponding line bundle, which form the space of the induced representation $ind^G_H(\chi_f)$. Its spectrum can be computed using the Frobenius reciprocity and Kostant's multiplicity formula, the details can be found in \cite{Kemp}.

\subsection*{Acknowledgements}
We are grateful to Ivailo Mladenov, Alex Strohmaier and especially to Alexey Bolsinov for useful discussions.


\begin{thebibliography}{99}
\footnotesize\itemsep=0pt

\bibitem{BJ}
Bolsinov A.V., Jovanovic B.{\it Magnetic flows on homogeneous spaces.} Comm. Math. Helv {\bf 83} (2008), 3, 679-700

\bibitem{Bott}
Bott R. {\it On induced representations.} In The mathematical heritage of Hermann Weyl (Durham, NC, 1987), 1-13, Proc. Sympos. Pure Math., 48, Amer. Math. Soc., Providence, RI, 1988.


\bibitem{Col}
Coleman S. {\it The magnetic monopole 50 years later.} In The Unity of the Fundamental Interactions (1983), A. Zichichi, editor.

\bibitem{Dirac}
Dirac P.A.M.
{\it Quantised singularities in the electromagnetic field.}
Proc. Roy. Soc. {\bf A 133} (1931), 60--72. 

\bibitem{E}
Efimov D. I. {\it The magnetic geodesic flows on a homogeneous symplectic manifold.} Sib
erian Mathematical Journal {\bf 46, no.1} (2005), 83-93.

\bibitem{FV}
Ferapontov E.V., Veselov A.P. 
{\it Integrable Schr\"odinger operators with magnetic fields: Factorization method on curved surfaces.}
J. Math. Phys. {\bf Vol. 42, No. 2} (2001), 590--607.

\bibitem{Fierz}
Fierz M. {\it On the theory of particles with magnetic charge.} Helv. Phys. Acta {\bf 17} (1944), 27-34.

\bibitem{FH}
Fulton W., Harris J. {\it Representation Theory}. Springer, 1991.

\bibitem{Hu} Hurt N.E., 
\emph{Geometric Quantization in Action},  Mathematics and its Applications; 8,  (1983), D. Reidel Publishing Company.

\bibitem{Kemp}
Kemp G.M. {\it Algebra and geometry of Dirac's magnetic monopole.} PhD Thesis, Loughborough University, June 2013.

\bibitem{Kir}
Kirillov, A. A. {\it Geometric quantization.} Dynamical systems IV, Encyclopaedia Math. Sci., 4, Springer, Berlin, 2001.

\bibitem{Kir1}
Kirillov A.A.,
Lectures on the orbit method, \textit{Graduate Studies in Mathematics}, American Mathematical Society, \textbf{Vol. 64} 2004.

\bibitem{KN}
Kobayashi S., Nomizu K. {\it Foundations of Differential Geometry.} Vol. I.  John Wiley and Sons, New York-London, 1963 

\bibitem{Mlad}
Mladenov I. M., Tsanov V. V. {\it Geometric quantisation of the MIC-Kepler problem.} 
J. Phys. A20 (1987), no. 17, 5865-5871.

\bibitem{NS}
Novikov S.P., Schmelzer I.
{\it Periodic solutions of Kirchhoff's equations for the free motion of a rigid body in a fluid and the extended theory of Lyusternik - Shnirel'man - Morse (LSM). I.}
Funct. Anal. Appl. {\bf Vol. 15, No. 3} (1981), 54--66.

\bibitem{Prieto}
Tejero Prieto C. {\it Quantization and spectral geometry of a rigid body in a magnetic monopole field.} Differential Geom. Appl. {\bf 14} (2001), no. 2, 157-179.

\bibitem{Rei}
Reiman A.L. {\it Relativistic and Galilean-invariant classical mechanical systems.} In Differential Geometry, Lie Groups and Mechanics. Zapiski Nauch. Seminarov LOMI. Vol. 37 (1973), 47-52 (in Russian).

\bibitem{Sch} 
Schr\"odinger E. {\it A method of determining quantum mechanical eigenvalues and eigenfunctions.} 
Proc. Royal Irish Acad. {\bf A 46} 9--16 (1940); {\it Further studies on solving eigenvalue problems by factorisation} , ibid. 183--206 (1941).

\bibitem{Tamm} Tamm I.
{\it Die verallgemeinerten Kugelfunktionen und die
 Wellenfunktionen eines Elektrons in Felde eines Magnetpoles.}
 Zeitschrift f\"ur Physik, {\bf 71}, 141--150 (1931).

\bibitem{WY}
Wu T.T, Yang C.N.,
{\it Dirac monopole without strings: monopole harmonics.}
Nuclear Physics B {\bf 107} (1976), 365--380. 









\end{thebibliography}
\end{document}